# Risk Quantification Associated with Wind Energy Intermittency in California

Sam O. George, *Member, IEEE*, H. Bola George, Ph.D. and Scott V. Nguyen, Ph.D.

*Abstract*--As compared to load demand, frequent wind energy intermittencies produce large short-term (sub 1-hr to 3-hr) deficits (and surpluses) in the energy supply. These intermittent deficits pose systemic and structural risks that will likely lead to energy deficits that have significant reliability implications for energy system operators and consumers. This work provides a toolset to help policy makers quantify these first-order risks. The thinking methodology / framework shows that increasing wind energy penetration significantly increases the risk of loss in California. In addition, the work presents holistic risk tables as a general innovation to help decision makers quickly grasp the full impact of risk.

*Index Terms*--California, renewable energy, risk analysis, systems engineering, wind power generation.

## I. INTRODUCTION

THIS work is presented as a companion to our paper submitted for publication [1]. Two important outputs in that report are: (1) If the components in California's Renewable Portfolio Standard (RPS) grow at current rates, wind energy will constitute 15% of the state's energy generation by 2016;[1] (2) The state has energy reserve capacity between 2 and 5 GWh (5 to 10% of total 2009 energy demand) consisting of spinning (and other) reserves. For the wind component of the RPS (wRPS) greater than 5%, the current reserve capacity is too low and not correctly configured to mitigate the risks associated with wind intermittency.

The random, frequent (hour-to-hour) and large changes in wind energy output create deficits (and surpluses) (Fig. 1) that impose new stresses and risks for the stability of the electric grid infrastructure. Without utility-scale energy storage assets, the nature of these risks is significantly different from other conventional energy sources like fossil fuels.

Wind energy intermittencies create systemic and structural risks. In this context, "systemic risk" defines risk that is tied to the hour-to-hour operation of the energy grid. This type of risk affects the entire grid or major segments of it on a dynamic basis. Structural risk is that associated with chronic shortfalls due to insufficient energy generation. This is a strategic planning problem that may stem from current use of simplistic macro-exchange equations in which annualized average energy from wRPS sources is made equivalent to energy produced from other non-renewable sources. Since the state does not plan to install redundant non-renewable generating equipment to compensate for the intermittencies of wind energy, the systemic / structural risks will rise as the fraction of wRPS increases. In this scenario, beyond about 5% wind penetration [1] the state may experience risks leading to losses of tens of billions of dollars.

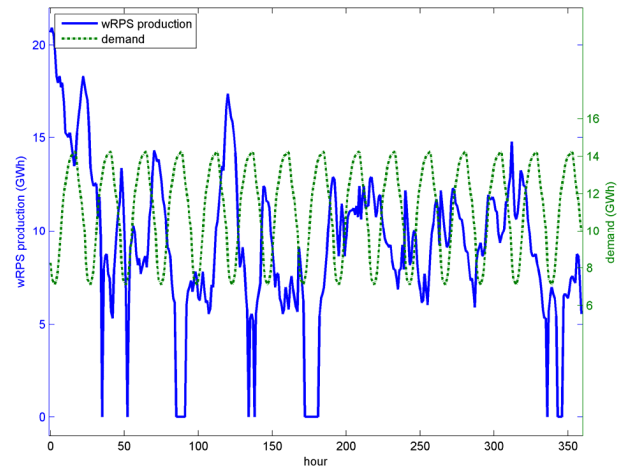

Fig. 1: wRPS = 15% energy production and load demand profiles vs. hour for a scenario containing 15 contiguous days. No energy is produced when wind speed falls below 4 m/s or exceeds 25 m/s. In this case, no energy is produced in 26 (7.22%) out of 360 hours.

In this work, the focus of wind energy risk planning is energy stability—not safety, as is more common in nuclear energy. Under normal conditions (i.e., no storm or excessive load demand), it is not possible to forecast wind energy output with a high degree of confidence. As shown with the application of the hour-to-hour auto-correlation function (hhACF) in [1], wind energy has large short-term predictive uncertainty. These, coupled with the fact that wind energy generation may fall to zero, are the basic factors of energy instability represented by wind.

This work provides a toolset for policy makers struggling to make the right energy policy choices that will have profound multi-decades impact. In our view, proactive RPS energy policy choices must be balanced with appropriate understanding and mitigation of systemic and structural risk. The consequence of inadequate risk strategies possibly exposes the state to energy deficit crises in 6 to 10 years.

Why should Californians take this seriously? There is precedence of energy instability in our recent past; in the

Financial support for this work is provided by GridByte, Inc., *Energy Policy Analytics* Practice.

S. O. George and H. B. George, Ph.D. are with GridByte, Inc., 65 Enterprise, Aliso Viejo, CA 92656 USA (e-mail: info@gridbyte.com).

S. V. Nguyen, Ph.D. is with Shell Projects and Technology, Innovation and R&D Division, Houston, TX 77002 USA.

[1] The quantity of 15% by 2016 is an illustrative benchmark from the data extrapolation presented in [1].

structural energy crises of 2000-2001, it is estimated that California lost $40 to $45 billion (about 3.5% of Gross State Product (GSP)) [2]. During this period, the state experienced rolling blackouts (load shedding) over 38 days [3] as energy demand exceeded supply by an average of 600 MW. In some cases, electricity customers lost power for up to 16 hours.[2] Again, as cited in [1], notable recent precedents exist in Denmark and Texas.

A note about reading this document: The purpose is to provide an analytics framework for energy risk quantification. Of course, it is possible that businesses and government will not stand by and allow wRPS risks to become chronic. The reality is that we are operating under mandates codified in California law (CA AB 32 / Governor's executive order) to achieve 33% RPS (RPS33) by 2030. The logical action is that risk mitigation infrastructure will be added to cope with the inherent intermittencies of wind energy. One essential component of risk mitigation infrastructure may include utility-scale storage.

## II. WIND ENERGY RISK ANALYTICS

### A. The Faulty Energy-Exchange Macro Equation

Without significant utility-scale storage, wind energy should not be equated with energy from conventional sources (e.g., fossil-based). The underlying risk is that wind energy has large random short-term (sub 1-hr to 3-hr) intermittencies, as shown in Fig. 1, that necessitate constant compensation [1]. The energy-exchange macro equations equate the statistical average energy from wRPS generator sites to the absolute energy produced from conventional sources. Moreover, the statistical averages are often derived from data measured in annual terms. This type of averaging masks the short-term intermittencies that are important for grid stability. So, effectively, there are two or more[3] levels of averaging that lead to a faulty outcome.

Wind energy dispatch is a real-time scheduling problem. 1 GWh of energy from a fossil fired plant $\neq$ 1 GWh of energy from any wind farm (or collection of wind farms). To make the macro energy-exchange equation "work" today, California relies on interruptible power agreements with large energy consumers [4]—this so-called demand-side compensation, is another major element of risk. As shown by the precedence of California's 2000-2001 energy crises, large businesses that use these interruptible power contracts are much less tolerant of blackouts. During the energy crises, as exemplified by Fruit Growers Supply Co., a number of companies either sought to extricate themselves from these contracts [5] (see Section II-G) or chose to keep power flowing at expensive premium rates. If we use precedence as a guide, then the reliance on interruptible power contracts poses a serious risk to grid stability.

### B. Definition of Risk as used in this work

Formally, risk, $R_\$$, is the product of two components: (The probability of an Energy Deficit, $P_{E_{Deficit}}$) × (The Impact of Energy Deficits, $I_{E_{Deficit}}$).

$$R_\$ = P_{E_{Deficit}} \times I_{E_{Deficit}} \qquad (1)$$

It is our goal to present solutions that comply with equation (1). When not possible, we will use a set of heuristics that follow the spirit of the equation.

### C. Definition of Energy Deficit

Measured on a short-term basis (sub 1-hr to 3-hr), energy deficit ($E_{Deficit}$) is the difference between demand load ($E_{Demand}$) and wRPS generation output ($E_{wRPS}$) + some reserve capacity ($E_{ResCap}$).

$$E_{Deficit} = E_{Demand} - E_{wRPS} + E_{ResCap} \qquad (2)$$

We assume that the 2 to 5 GWh spinning, and other, reserves can produce the fast response [1] required to compensate for the short-term intermittencies in wind energy output. Implicit in this assumption is the requirement that the fast compensation reserves must have non-stochastic real-time stability as compared to $E_{wRPS}$ output.

### D. Risk Factors

While we restrict our discussion to the first-order risk associated with energy deficits, it is important to note that the dynamics of wRPS integration create a number of additional and significant risk factors. Table 1 presents a partial summary of risk factors that are considered in the context of this work.

### E. Probability of n-hr Deficit Clusters

wRPS profiles, such as in Fig. 1, contain $n$-hr "natural" clusters of energy deficit, i.e., contiguous hour-to-hour deficits. The cluster lengths may be one or multiple hours long. For example, from Fig. 1, the "natural" formation produces clusters ranging in length from 1 hour to 15 hours. While natural clusters are good for description of the experimental data, they pose a challenge for forming reliable probability metrics. Specifically, it is difficult to talk systematically about the probability of "naturally" formed clusters of different sizes.

In the models for this work, we use "synthetic" clustering. We use two different counting methods that bound the range of probabilities of synthetic $n$-hr cluster sizes. These synthetic clusters are deliberate constructs to ensure that the probability computations yield consistent results.

Application of these probability counting methods (described in *Section II-F*) produces probabilities for synthetic $n$-hr windows; i.e., 1 hr, 2 hr, 3 hr, ... etc. The important point is that the probabilities associated with the synthetic $n$-hr clusters exist orthogonally; i.e., they can all be combined in the same space without affecting the accuracy of the overall probability estimate. The orthogonal properties allow

---

[2] Granted, the causes of the 2000-2001 energy crises stem from the implementation of California's electricity industry deregulation efforts—a scenario that is different from wRPS implementation. However, the crises of 2000-2001 are instructive in the simple point that they should not have happened. If these crises fell outside the projected normative behavior—then their occurrence is instructive for wRPS implementation as it shows that outlier events can and will occur with serious consequences. But importantly, the hour-to-hour systemic risk may become the normative as wRPS penetration increases.
[3] There is one implicit level of averaging in that reported wind speeds are averages derived from Weibull probability plots / analyses.



application of different cost impact factors as discussed in *Section II-G*.

Fig. 2 presents a sample application of the synthetic deficit cluster probability algorithm (SDCPA) described in *Section II-F*. In this example, we utilize the wind energy generation scenario shown in Fig. 1 as the basis. The reserve capacity of 5 GWh is assumed to be readily dispatchable to compensate for any / all wind intermittencies.

### F. Synthetic deficit cluster probability algorithm (SDCPA)

This section presents the synthetic deficit cluster probability algorithm (SDCPA). The algorithm is presented in pseudo-code for simplicity. The SDCPA uses two methods to calculate the probability of energy deficit clusters. Method 0 counts all non-overlapping $n$-hr synthetic energy deficit clusters. This method produces a slight undercounting as cluster sizes increase. Method 1 counts overlapping $n$-hr synthetic clusters. This method produces a slight over-count. The SDCPA is implemented as follows:

//Let ...

| | | |
|---|---|---|
| $e_{Demand}$ | = vector | → energy demand profile by hour |
| $e_{wRPS}$ | = vector | → wRPS energy production profile by hour |
| $e_{Deficit}$ | = vector | → containing energy deficits |
| $v$ | = vector | → binary thresholds |
| $c_0$ | | → representing clusters (method 0) |
| $c_1$ | = vector | → representing clusters (method 1) |
| $E_{ResCap}$ | = scalar | → total reserve capacity in grid |
| $N$ | = scalar | → number of hours in $e_{Demand}$ and $e_{wRPS}$ profiles |
| $h$ | = scalar | → subscript for hour |
| $i$ | = scalar | → subscript for clusters |
| $n_0$ | = vector | → Total number of clusters and non-clusters; method 0 |
| $n_1$ | = vector | → Total number of clusters and non-clusters; method 1 |
| $p_0$ | = vector | → probability of clusters; method 0 |
| $p_1$ | = vector | → probability of clusters, method 1 |

//Create deficit vector ...
For ($h = 1$ to $N$, $h$++) {

$e_{Deficit}[h] = (e_{Demand}[h] - e_{wRPS}[h] - E_{ResCap}) \times$
$\qquad [(e_{Demand}[h] - e_{wRPS}[h]) \geq 0]$;

$v[h] = e_{Deficit}[h] > 0$; // $v[h]$ is a 1D vector of 0's or 1's
}

// Create cluster vectors (method 0 and method 1) ...
For ($i = 1$ to 23; $i$++) { // step size is always 1
    For ($h = i$ to $N$; $h$++) {
        $r = h - i + 1$;

$\qquad c_0[i]$ += $\prod_{j=r}^{h} v[j]$;

$\qquad n_0[i]$++;
    }
    $m = 0$;
    For ($h = i$ to $N$; $h + m$) { // step size is 1 or $i$
        $q = h - i + 1$;

$\qquad c_1[i]$ += $\prod_{j=q}^{h} v[j]$;

$\qquad m = i \times \prod_{j=q}^{h} v[j]$;

        if ($m == 0$, $m = 1$)
            $n_1[i]$++;
    }

// Calculate probabilities ...

---

Table 1: wRPS Risk Factors Matrix

| Risk Factor | Primary Risk Factors | Notes |
|---|---|---|
| 1 | Large-magnitude hour-to-hour intermittencies in wind energy generation capacity. | With the state's current reserve capacity and beyond 5% wind penetration [1], the large intermittencies of wind energy may lead to deficits of unpredictable magnitude and duration. |
| 2 | Rate of change of wind energy generation output. | Large hour-to-hour changes in wind energy generation create ramp-rate problems. Because it is difficult to predict the short-term output of wind farms in the grid, it may not be possible to deliver compensation energy to grid-segments when required. The ramp-rate problem requires fast-response generators to compensate for sudden changes in wind energy output. From Fig. 1, there are several instances in which the hour-to-hour wind energy output varies by more than 10 GW. For example, a 10 GW deficit requires the fast compensation generators to produce a sustained ~167 MW/min. Ramp-rates of this order are difficult to sustain with the mix of generating assets currently deployed in California. In our estimate, such a ramp rate requires tens of fast compensation generators. |
| 3 | Random periods of wind energy deficits and surpluses. Random intervals between wind energy deficits / surpluses. | The length of time associated with wind energy deficits (or surpluses) is random. This produces large-scale planning uncertainty for grid operators. If these deficits translate to blackouts, the effects on the economy would magnify non-linearly. For example, the grid operators may need to maintain compensation generators in sub-optimal standby mode for many more hours than required. This practice constitutes a large loss of revenue. |
| 4 | Transmission constraints. | Compensating for wind energy deficits is constrained by the transmission infrastructure. Even if all the compensating generation capacity is dispatchable, there is a real possibility that the energy may not get to the clients because of transmission bottlenecks. |
| 5. | Rate of Implementation of wRPS. | Evolution of the current grid follows an innovation trajectory spanning 100 to 150 years. Most utility-scale wRPS implementation experience is less than 20 year. The lack of field and technical experience / data constitutes an element of risk. |



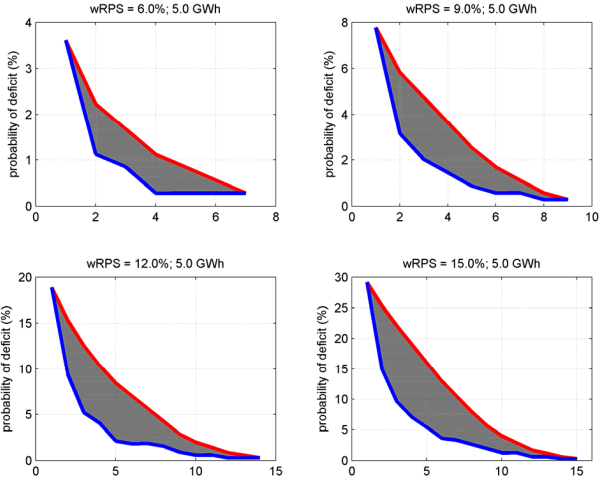

Fig. 2: Probability of energy deficits at different wRPS penetration levels with 5 GWh reserve capacity vs. *n*-hr deficit clusters in both overlap (red curve) and non-overlap (blue) formulations.

$$p_0[i] = \frac{c_0[i]}{n_0[i]};$$

$$p_1[i] = \frac{c_1[i]}{n_1[i]};$$

}

### G. Impact of n-hr Wind Energy Deficits

The second component of risk (eq. 1) is the dollar-impact ($I_{E_{Deficit}}$) associated with *n*-hr energy deficits. Impact is the product of the normalized loss-per-hour ($L_{hr}$) and the total number of hours in the corresponding *n*-hr deficit clusters ($N_{n-hr}$).

$$I_{E_{Deficit}} = L_{hr} \times N_{n-hr} \qquad (3)$$

Growth in the fraction of energy from wRPS means that most of California's highly interconnected economy will be adversely affected by energy deficits resulting from wind intermittencies. Thus we calibrate $L_{hr}$ based on broad application of the sector customer damage functions (SCDF) developed in [6]. $L_{hr}$ is based on the weighted average cost of 1-hour energy interruption for all sectors of the economy. $L_{hr}$ uses a loss basis of \$8.76 / kWh in 1996 dollars for the entire United States. In this work, our illustrations are computed with the 1996 loss basis of \$8.76 / kWh.[4] As noted in [6][7], the economic losses associated with multi-hour deficits is non-linear, in that the "interruption costs increase with duration in a non-linear manner." Further, the random and large intermittencies of wRPS generation complicate the calculations. To simplify our models, we apply the 1-hour loss-basis linearly across all clusters of deficits. Thus, the cost of larger deficit clusters are underestimates.

Let us posit that California's Gross State Product (GSP) is at parity when there is energy stability; i.e., demand is equal to generation on an hour-to-hour basis. Parity means that the grid, based on data for the entire United States, can supply, 100% energy with a reliability of 99.96% (corresponding to the highest reliability of 3.5 hours of blackouts per year) [6].

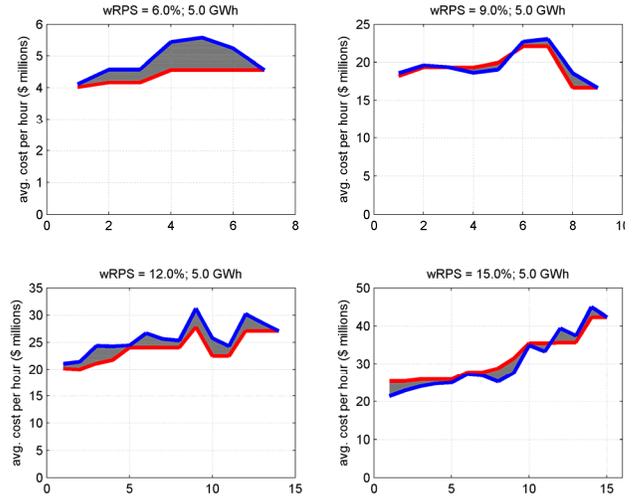

Fig. 3: Dollar-impact ($L_{hr}$) of energy deficits at different wRPS penetration levels with 5 GWh reserve capacity vs. *n*-hr deficit-clusters. $L_{hr}$ is based on a loss-basis of \$8.76 / kWh.

Using $L_{hr}$ and the SDCPA in *Section II-F*, the model computes the 1-hour dollar-impact as shown in Fig. 3.

For example, Fig. 3 shows $L_{hr}$ at four wRPS penetration levels vs. synthetic cluster sizes based on a reserve capacity of 5 GWh. At wRPS = 15%, the $L_{hr}$ hour ranges from \$21.38 million to \$43.32 million/hr.

From application of the synthetic deficit cluster probability algorithm in Section *II-F*, we obtain the total number of hours corresponding to each *n*-hr cluster shown in Fig. 4.

For perspective, it is useful to review the dollar-impact of energy deficits during the 2000-2001 energy crises on one California farming operation [5][8]. As shown, losses mount and multiply quickly during multi-hour blackouts. As further illustrated in [7], the losses are often under-reported. Businesses generally have low tolerance for blackouts—they quickly begin to invest in blackout mitigation equipment (e.g., backup generators). These investments are non-incremental business-continuity insurance expenditures that, in addition to maintenance, represent loss of profit. The experience of the

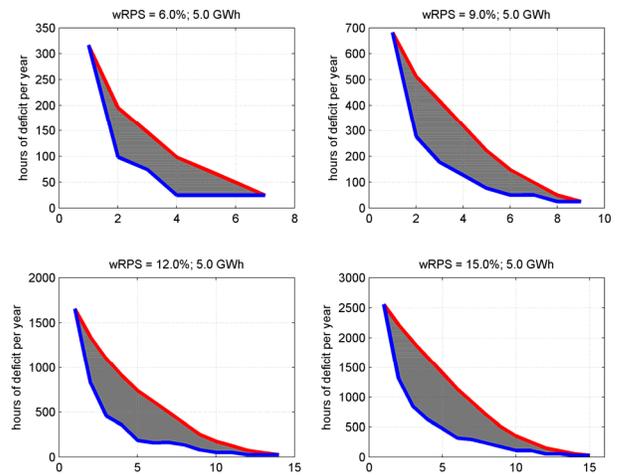

Fig. 4: Summation of total number of hours corresponding to each n-hr cluster of energy deficits vs. *n*-hr clusters. The reserve capacity is 5 GWh.

---

[4] If we correct for inflation and other economic factors, the loss-basis is about \$16.08 / kWh in 2009 dollars. The estimates in [6] represent a comprehensive loss basis from which other estimates can easily be drawn.

Land O'Lakes cooperative also illustrates that reliance on interruptible power contracts is not workable if the energy deficits become frequent. Thus, the fact that large businesses sign up for interruptible power programs should not be regarded as an indication of high risk tolerance. Rather, it is an exercise in business operating-cost minimization because these programs offer significant discounts for participation.

### H. Risk Associated with Energy Deficits

As defined in eq. 1 (*Section II-B*), the annualized Risk is the product of the probability of deficits (*Section II-E*) and the dollar-impact of these deficits (*Section II-G*). Building on the examples in these sections, we present Fig. 5—a view of the risk associated with synthetic clusters of energy deficits. Fig. 5 shows how risk grows with increasing wRPS penetration.

To present a holistic view of risk, we show two snapshots of risk tables in *Section II-J*.

### I. How wRPS intermittency reduces Reliability

To the first order, the risk posed by wRPS intermittencies changes California's reliability expectations significantly. Reliability is one minus the probability of energy deficits. At wRPS = 15% and reserve capacity of 5 GWh, California's energy generation reliability may drop to 70.83%. This, as compared to nominal baseline performance of 99.96%, represents many hundred hours of energy deficits. Fig. 6 presents the reliability profile vs. wRPS penetration levels for the generation / demand profiles in Fig. 1 for 1-hour deficits. Beyond the first order, there are higher order risks associated with the frequency of energy deficits; i.e., in the wRPS scenario, the deficits occur more frequently and randomly. The associated loss of such instability is largely unknown, but potentially as large as the first order risk presented in this work.

### J. Risk Tables—A classical view

To present a comprehensive view of risk, this section utilizes a classical method similar to that from the actuarial sciences. The underlying equations are generally unwieldy—thus the tabular format is more accessible. In this section we present two examples in Table 2 (wRPS = 6%) and Table 3 (wRPS = 15%). From Table 2, California's grid can 'sustain' wRPS = 6% with 5 GWh reserve capacity to produce associated risk between 0 and $1 billion. In contrast, Table 3 shows that the risk associated with wRPS=15% exceeds $50 billion at a reserve capacity of 5 GWh (10% of peak demand). The risk profile is much less for a reserve capacity of 10 GWh (20% of peak demand)—but as discussed in Section III, California has to re-evaluate the opportunity cost of wRPS vs. deployment of 10 GWh reserve capacity.

### III. Conclusion / Solutions

In addition to the conclusions in [1], this work shows how / why risk quantification analytics methodology should be included in California's wind RPS strategy. The risk tables in this work provide a holistic insight into the probable losses associated with various wRPS penetration levels and reserve capacities. The loss-basis shown is conservative. This is further calibrated against estimated losses from California's 2000-2001 energy crises. Using this loss-basis, we estimate that California's risk exposure in a wRPS = 15% scenario could exceed tens of billions per year in 2009 dollars.

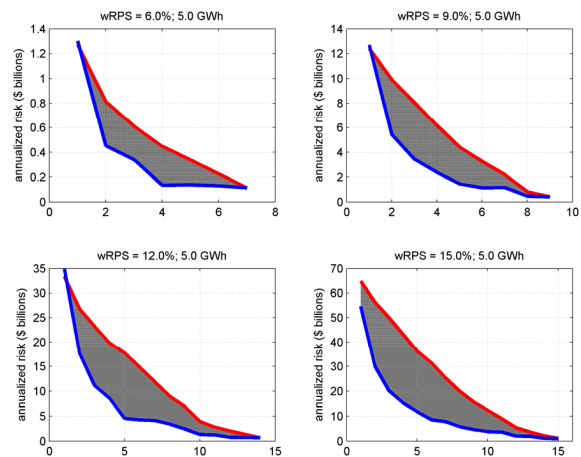

Fig. 5: Conservative annualized risk vs. *n*-hr deficit cluster sizes for the state of California at various wRPS penetration levels. The reserve capacity is 5 GWh.

There is no easy way out—at high wRPS penetration levels (e.g., 15%) a significant fraction of the state's GSP [9] will be lost directly from wRPS intermittency. If not directly, the loss in GSP will be felt indirectly as non-incremental expenditures are diverted to wRPS intermittency risk mitigation.

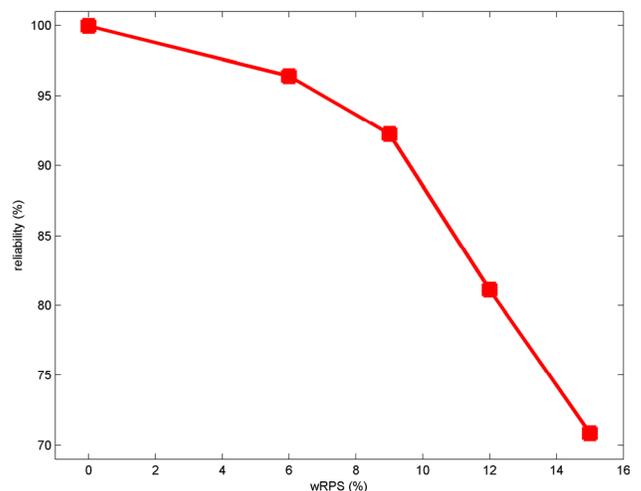

Fig. 6: Energy-supply reliability vs. wRPS penetration levels for the generation / demand profiles shown in Fig. 1. At wRPS = 15% and reserve capacity = 5 GWh, the reliability drops to 70.83%.

Table 2: Annualized Risk (in 2009 dollars) at various wRPS = 6% penetration vs. *n*-hr energy deficit clusters
(The reserve capacities are 2 and 5 GWh)

| wRPS (%) | Reserve Capacity (GWh) | *n*-hr Cluster Size (hours) | Method 1: Non-overlap method | | | | | Method 0: Overlap Method | | | | |
|---|---|---|---|---|---|---|---|---|---|---|---|---|
| | | | Probability of n-hr Energy Deficit (%) | Total Energy Deficit Hours per Year (hours) | Total Energy Deficit Per Year (GWh) | Economic Loss per Energy Deficit Hour ($MM) | Risk per Year ($BB) | Probability of n-hr Energy Deficit (%) | Total Energy Deficit Hours per Year (hours) | Total Energy Deficit Per Year (GWh) | Economic Loss per Energy Deficit Hour ($MM) | Risk per Year ($BB) |
| 6 | 2 | 1 | 29.2 | 2556.8 | 68.36 | 8.55 | **21.860** | 29.2 | 2556.8 | 119.94 | 10.144 | **25.934** |
| 6 | 2 | 2 | 15.0 | 1316.3 | 65.94 | 9.149 | **12.043** | 25.3 | 2222 | 117.68 | 10.146 | **22.544** |
| 6 | 2 | 3 | 9.6 | 841.8 | 64.81 | 9.592 | **8.074** | 22.1 | 1934.4 | 117.54 | 10.334 | **19.990** |
| 6 | 2 | 4 | 7.1 | 619.8 | 62.26 | 9.872 | **6.119** | 19 | 1669.7 | 117.54 | 10.334 | **17.255** |
| 6 | 2 | 5 | 5.4 | 473.8 | 61.79 | 9.976 | **4.727** | 16 | 1403.5 | 117.54 | 10.334 | **14.504** |
| 6 | 2 | 6 | 3.6 | 316.2 | 58.73 | 10.865 | **3.435** | 13 | 1135.9 | 113.67 | 11.092 | **12.599** |
| 6 | 2 | 7 | 3.3 | 292.2 | 59.27 | 10.742 | **3.139** | 10.5 | 916.2 | 113.67 | 11.092 | **10.163** |
| 6 | 2 | 8 | 2.6 | 230.7 | 45.5 | 10.101 | **2.330** | 7.9 | 695.3 | 109.13 | 11.537 | **8.022** |
| 6 | 2 | 9 | 1.9 | 168.6 | 44.7 | 11.026 | **1.859** | 5.7 | 498.1 | 96.34 | 12.581 | **6.266** |
| 6 | 2 | 10 | 1.2 | 108.2 | 47.21 | 13.975 | **1.512** | 4 | 349.6 | 79.96 | 14.201 | **4.965** |
| 6 | 2 | 11 | 1.3 | 109.6 | 49.4 | 13.293 | **1.457** | 2.9 | 250.5 | 79.96 | 14.201 | **3.557** |
| 6 | 2 | 12 | 0.6 | 51.9 | 42.64 | 15.778 | **0.818** | 1.7 | 150.7 | 44.98 | 14.266 | **2.150** |
| 6 | 2 | 13 | 0.6 | 52.2 | 43.85 | 14.977 | **0.781** | 1.1 | 100.8 | 44.98 | 14.266 | **1.437** |
| 6 | 2 | 14 | 0.3 | 25.3 | 28.37 | 17.993 | **0.455** | 0.6 | 50.5 | 28.6 | 16.93 | **0.855** |
| 6 | 2 | 15 | 0.3 | 25.3 | 28.6 | 16.93 | **0.429** | 0.3 | 25.3 | 28.6 | 16.93 | **0.429** |
| 6 | 5 | 1 | 3.6 | 316.6 | 5.09 | 4.111 | **1.301** | 3.6 | 316.6 | 5.88 | 4.017 | **1.272** |
| 6 | 5 | 2 | 1.1 | 98.5 | 4.12 | 4.576 | **0.451** | 2.2 | 195.3 | 4.69 | 4.166 | **0.814** |
| 6 | 5 | 3 | 0.8 | 74.3 | 4.64 | 4.577 | **0.340** | 1.7 | 146.9 | 4.69 | 4.166 | **0.612** |
| 6 | 5 | 4 | 0.3 | 24.6 | 2.45 | 5.44 | **0.134** | 1.1 | 98.2 | 3.59 | 4.559 | **0.448** |
| 6 | 5 | 5 | 0.3 | 24.6 | 3.14 | 5.579 | **0.137** | 0.8 | 73.9 | 3.59 | 4.559 | **0.337** |
| 6 | 5 | 6 | 0.3 | 24.7 | 3.54 | 5.242 | **0.129** | 0.6 | 49.4 | 3.59 | 4.559 | **0.225** |
| 6 | 5 | 7 | 0.3 | 24.8 | 3.59 | 4.559 | **0.113** | 0.3 | 24.8 | 3.59 | 4.559 | **0.113** |

With a holistic view of risk, the state needs to re-evaluate the opportunity costs associated with wRPS implementation. For example, one way to achieve wRPS = 15% is to invest in appropriate fast-response energy reserve capacity (such as combined-cycle gas-fired plants) or utility-scale storage assets [1].

The state must also re-examine whether reliance on interruptible power contracts as a means for maintaining grid stability is workable in the wRPS = 15% scenario. With the precedence of the 2000-2001 crises coupled with these large risks, it is our view that interruptible power contracts are not workable as wRPS energy deficits increase in frequency, randomness and length.

Table 3: Annualized Risk (in 2009 dollars) at various wRPS = 15% penetration vs. *n*-hr energy deficit clusters
(The reserve capacities are 2, 5 and 10 GWh)

| wRPS (%) | Reserve Capacity (GWh) | *n*-hr Cluster Size (hours) | Method 1: Non-overlap method | | | | | Method 0: Overlap Method | | | | |
|---|---|---|---|---|---|---|---|---|---|---|---|---|
| | | | Probability of *n*-hr Energy Deficit (%) | Total Energy Deficit Hours per Year (hours) | Total Energy Deficit Per Year (GWh) | Economic Loss per Energy Deficit Hour ($MM) | Risk per Year ($BB) | Probability of *n*-hr Energy Deficit (%) | Total Energy Deficit Hours per Year (hours) | Total Energy Deficit Per Year (GWh) | Economic Loss per Energy Deficit Hour ($MM) | Risk per Year ($BB) |
| 15 | 2 | 1 | 49.7 | 4358.6 | 440.48 | 35.239 | 153.593 | 49.7 | 4358.6 | 737.87 | 36.605 | 159.547 |
| 15 | 2 | 2 | 31.4 | 2751.4 | 435.32 | 36.469 | 100.339 | 45.1 | 3955.7 | 736.91 | 36.971 | 146.244 |
| 15 | 2 | 3 | 21.4 | 1878.4 | 415.43 | 37.263 | 69.996 | 41.1 | 3599.4 | 731.49 | 37.118 | 133.605 |
| 15 | 2 | 4 | 16.0 | 1406.9 | 394.08 | 38.037 | 53.514 | 37.3 | 3265.8 | 731.49 | 37.118 | 121.219 |
| 15 | 2 | 5 | 13.1 | 1151.5 | 404.58 | 37.817 | 43.545 | 33.4 | 2930.2 | 731.49 | 37.118 | 108.764 |
| 15 | 2 | 6 | 9.4 | 822.9 | 366.17 | 38.709 | 31.855 | 29.6 | 2592.8 | 727.22 | 37.986 | 98.490 |
| 15 | 2 | 7 | 7.7 | 677.0 | 383.82 | 40.575 | 27.471 | 26 | 2278.2 | 727.22 | 37.986 | 86.539 |
| 15 | 2 | 8 | 6.5 | 565.5 | 331.61 | 40.898 | 23.130 | 22.4 | 1961.8 | 715.81 | 38.996 | 76.503 |
| 15 | 2 | 9 | 5.1 | 445.1 | 315.56 | 38.919 | 17.325 | 19 | 1668.5 | 715.81 | 38.996 | 65.066 |
| 15 | 2 | 10 | 4.8 | 417.4 | 321.22 | 40.749 | 17.010 | 15.7 | 1373.6 | 715.81 | 38.996 | 53.565 |
| 15 | 2 | 11 | 4.4 | 385.7 | 321.89 | 43.309 | 16.704 | 12.3 | 1077 | 686.18 | 39.825 | 42.891 |
| 15 | 2 | 12 | 3.4 | 302.3 | 304.32 | 45.039 | 13.614 | 9.2 | 803.8 | 600.33 | 40.694 | 32.708 |
| 15 | 2 | 13 | 2.1 | 182.6 | 295.15 | 50.402 | 9.205 | 6.6 | 579.4 | 498.3 | 46.578 | 26.986 |
| 15 | 2 | 14 | 2.1 | 186.5 | 312.96 | 49.627 | 9.256 | 4.9 | 429.5 | 498.3 | 46.578 | 20.003 |
| 15 | 2 | 15 | 1.7 | 151.1 | 273.8 | 54.029 | 8.166 | 3.2 | 278.7 | 452.82 | 49.643 | 13.835 |
| 15 | 2 | 16 | 1.3 | 116.9 | 201.79 | 55.997 | 6.545 | 1.7 | 152.5 | 378.19 | 50.884 | 7.757 |
| 15 | 2 | 17 | 0.3 | 25.5 | 85.1 | 44.451 | 1.133 | 0.6 | 51 | 87.56 | 43.194 | 2.201 |
| 15 | 2 | 18 | 0.3 | 25.6 | 87.56 | 43.194 | 1.104 | 0.3 | 25.6 | 87.56 | 43.194 | 1.104 |
| 15 | 5 | 1 | 29.2 | 2556.8 | 170.91 | 21.375 | 54.651 | 29.2 | 2556.8 | 299.85 | 25.359 | 64.836 |
| 15 | 5 | 2 | 15.0 | 1316.3 | 164.85 | 22.874 | 30.109 | 25.3 | 2222 | 294.2 | 25.364 | 56.359 |
| 15 | 5 | 3 | 9.6 | 841.8 | 162.03 | 23.981 | 20.186 | 22.1 | 1934.4 | 293.84 | 25.835 | 49.974 |
| 15 | 5 | 4 | 7.1 | 619.8 | 155.65 | 24.681 | 15.298 | 19 | 1669.7 | 293.84 | 25.835 | 43.136 |
| 15 | 5 | 5 | 5.4 | 473.8 | 154.48 | 24.941 | 11.818 | 16 | 1403.5 | 293.84 | 25.835 | 36.260 |
| 15 | 5 | 6 | 3.6 | 316.2 | 146.83 | 27.163 | 8.588 | 13 | 1135.9 | 284.18 | 27.731 | 31.498 |
| 15 | 5 | 7 | 3.3 | 292.2 | 148.18 | 26.855 | 7.847 | 10.5 | 916.2 | 284.18 | 27.731 | 25.407 |
| 15 | 5 | 8 | 2.6 | 230.7 | 113.75 | 25.252 | 5.825 | 7.9 | 695.3 | 272.83 | 28.842 | 20.054 |
| 15 | 5 | 9 | 1.9 | 168.6 | 111.75 | 27.566 | 4.647 | 5.7 | 498.1 | 240.86 | 31.453 | 15.666 |
| 15 | 5 | 10 | 1.2 | 108.2 | 118.03 | 34.937 | 3.781 | 4 | 349.6 | 199.9 | 35.503 | 12.413 |
| 15 | 5 | 11 | 1.3 | 109.6 | 123.5 | 33.233 | 3.642 | 2.9 | 250.5 | 199.9 | 35.503 | 8.892 |
| 15 | 5 | 12 | 0.6 | 51.9 | 106.61 | 39.445 | 2.046 | 1.7 | 150.7 | 112.46 | 35.664 | 5.375 |
| 15 | 5 | 13 | 0.6 | 52.2 | 109.63 | 37.442 | 1.954 | 1.1 | 100.8 | 112.46 | 35.664 | 3.593 |
| 15 | 5 | 14 | 0.3 | 25.3 | 70.92 | 44.982 | 1.136 | 0.6 | 50.5 | 71.49 | 42.325 | 2.138 |
| 15 | 5 | 15 | 0.3 | 25.3 | 71.49 | 42.325 | 1.072 | 0.3 | 25.3 | 71.49 | 42.325 | 1.072 |
| 15 | 10 | 1 | 5.3 | 462.6 | 51.42 | 26.862 | 12.428 | 5.3 | 462.6 | 58.4 | 27.294 | 12.627 |
| 15 | 10 | 2 | 2.0 | 173.8 | 42.81 | 27.152 | 4.720 | 3.6 | 317.4 | 47.92 | 26.597 | 8.443 |
| 15 | 10 | 3 | 1.4 | 125.2 | 45.29 | 26.813 | 3.358 | 2.8 | 244.9 | 47.92 | 26.597 | 6.513 |
| 15 | 10 | 4 | 0.6 | 49.5 | 24.67 | 27.387 | 1.356 | 2 | 171.9 | 42.33 | 28.912 | 4.969 |
| 15 | 10 | 5 | 0.6 | 49.8 | 32.4 | 28.773 | 1.433 | 1.4 | 123.1 | 42.33 | 28.912 | 3.560 |
| 15 | 10 | 6 | 0.6 | 50.1 | 39.7 | 29.374 | 1.471 | 0.8 | 74.1 | 42.33 | 28.912 | 2.142 |
| 15 | 10 | 7 | 0.3 | 24.8 | 26.48 | 33.598 | 0.832 | 0.3 | 24.8 | 26.48 | 33.598 | 0.832 |

## V. Biographies

Sam O. George (BS 1993) is an EE graduate of Iowa State University, Ames, Iowa.

His competencies span a range of practices—high frequency / fidelity IC SoC product design / development (high frequency communications chipsets, power control subsystems and chipsets, multi-loop timing recovery designs, e.g., Fractional-N PLLs, data converters, et. al), BIST, systems and software architecture / modeling, SaaS, computer algorithms, Intellectual Property monetization and business problem solving / BPO.

Sam has led GridByte® for the past seven years, a multi-practice consultancy that, among other things, produces a range of risk-optimized strategic decision analytics tools / solutions for corporate and government clients. This innovation to management consulting combines engineering science and applied mathematics with systems expertise.

In addition to consulting positions at a number of fortune 500 companies, his career experience includes design management and IC lead positions at GlobespanVirata and Hughes Network Systems. In these positions he led high-frequency CMOS / BiCMOS chipset designs and had direct design responsibility for RF / mixed-signal / analog sub-systems, systems architecture, device modeling, yield optimization and DFM.

**H. Bola George** (Ph.D. 2007) is an applied physics graduate of Harvard University.

At GridByte, Inc., Bola has been focused on applying analytical frameworks to modeling of real-time physical systems, in particular, energy. Prior to employment at GridByte, Bola's work spanned investigation of surface mass transport mechanisms governing the formation of nanoscale features, materials development, and involvement with development of analytical tools to quantify morphological features.

**Scott V. Nguyen** (Ph.D. 2006) is a Senior Physicist in Shell's Innovation and R&D division where he identifies and develops technology applicable to the sustainable development of unconventional hydrocarbon resources, focusing on techno-economics, energy management, and greenhouse issues. He is a physics graduate of Harvard University and also currently serves on the advisory committee to the American Institute of Physics Corporate Associates.